\documentclass[a4paper,12pt,english]{article}
\usepackage{geometry}
\usepackage{float}
\usepackage{graphicx}
\usepackage{amsmath,amsthm}
\usepackage{enumerate}
\usepackage[american]{babel}
\usepackage[usenames, dvipsnames]{color}

\makeatletter
\newcommand{\doublespacing}{\let\CS=\@currsize\renewcommand{\baselinestretch}{1.2}\tiny\CS}

\date{}

\usepackage{xcolor}

\begin{document}
\doublespacing
\title{ Viscous Dark Energy and Mass-Varying Dark Matter in Lyra Manifold: Cosmological Dynamics and Observational Constraints}
\author{Giridhari Deogharia $^1$, Madhurima Pandey $^1$ and Ashadul Halder$^2$\footnote{Corresponding author: E-mail: ashadul.halder@gmail.com }}
\maketitle
\noindent \centerline{$^1$ School of Applied Science and Humanities,}
\centerline{Haldia Institute of Technology, Haldia-721657, W.B., India}
\noindent \centerline{$^2$ Department of Physics,}
\centerline{St. Xavier's College, 30, Mother Teresa Sarani, Kolkata-700016, India}

\maketitle

\noindent \underline{\hspace*{16.5cm}}\\
\setlength{\unitlength}{1cm} \textbf{Abstract:}
We investigate the cosmological dynamics of a universe described by Lyra's geometry in the presence of dark energy (DE) and dark matter (DM). Dark energy is modeled as a quintessence scalar field with bulk viscosity, while dark matter is allowed to interact with the scalar sector. The displacement vector field, arising naturally in Lyra's manifold, provides an additional geometric contribution. By employing dynamical system techniques, we analyze stability properties and late-time attractors. Our results indicate that viscosity and DE--DM interaction enrich the phase space structure and can help address both the cosmic acceleration and the coincidence problem. Furthermore, by performing a Markov Chain Monte Carlo (MCMC) analysis with recent observational datasets, we derive best-fit values for the model parameters that exhibit good consistency with current data.

\noindent \underline{\hspace*{16.5cm}}\\
\pagestyle{myheadings}
\bigskip

\section{Introduction}
Modern cosmology has undoubtedly entered into a data - driven era, propelled by a series of groundbreaking observations that have reshaped our understanding of the Universe's large - scale dynamics. Over the past two decades, diverse and independent probes, such as Type Ia supernovae (SNIa) \cite{Riess1998, Perlmutter1999}, baryon acoustic oscillations (BAO) \cite{bao} and cosmic microwave background radiation (CMBR) \cite{cmbr1,cmbr2} have provided compelling evidence for the late - time accelerated expansion of the Universe. Observational cosmology has established that the luminous baryonic component, comprising star, interstellar gas, and other radiating matter, accounts for only about 4.9 \% of the total mass - energy density of the Universe. High precision measurements from the PLANCK satellite have provided captivating evidence that the remaining cosmic inventory is dominated by two non - luminous constituents; such as dark matter (DM) and dark energy (DE). The DM component is non - baryonic and collisionless in nature. It's presence can be understood solely through gravitational effects on galactic rotation curves, gravitational lensing, large scale structure formation etc. Dark matter contributes 26.8 \% of the total energy density of the Universe. The rest around 69.3 \% is attributed to DE. It is a hypothetical form of energy, which exists negative pressure that acts in opposition to gravity and driving the accelerated expansion of the Universe observed in the late - time cosmological epoch. 

As discussed earlier, the negative pressure attributed to DE counteracts the attractive influence of gravity, thereby causing the accelerated expansion of the Universe and preventing an ultimate gravitational collapse. Therefore, a wide range of theoretical models have been developed to explain the physical origin and dynamical properties of DE and the ensuing late - time acceleration. The most straight forward and widely accepted explanation is the cosmological constant ($\Lambda$), introduced by Einstein in his field equations of general relativity as a constant vacuum energy density term. This forms the basis of the $\Lambda$ - cold dark matter (CDM) model, where DE is characterized by a constant equation of state parameter, $\omega = p/\rho = -1$, implying a uniform energy density through out spacetime. High - precision data from the Planck CMB mission have reaffirmed the $\Lambda$CDM model as the observationally consistent and theoretically feasible description of the Universe's large scale dynamics \cite{cmbr2}. 

Although the $\Lambda$CDM model remains one of the cornerstones of modern cosmology, it is not yet regarded as the ultimate framework for understanding our Universe \cite{Weinberg1989}. From a theoretical perspective, the underlying physics of DE and DM continues to elude a fundamental explanation. The model also suffers from persistent conceptual challenges, notably the fine tuning and cosmic coincidence problems associated with the cosmological constant \cite{Weinberg1989,Zlatev1999}. It has been observed that, while $\Lambda$ successfully reproduces the large scale distribution of matter and the evolution of cosmic structures, it encounters several inconsistencies on galactic and sub - galactic scales, such as the missing satellites, core - cusp and too - big - to fail problem \cite{bullock}. In parallel, the advent of high - precision cosmological observations has brought to light additional tensions within the $\Lambda$CDM framework. The most prominent one among these is the $H_0$ tension, referring to the nearly 5$\sigma$ discrepancy between the Hubble constant ($H_0$) inferred from Planck CMB measurements and that derived from local standard candle and gravitational lensing time delay observations \cite{late} - \cite{wong}. Another emerging inconsistency is the $S_8$ tension, which reflects 1 - 2$\sigma$ deviation between the parameter $S_8 = \sigma_8 (\Omega_m)^{0.5}$ obtained from Planck data and the results from weak lensing surveys \cite{s8} -\cite{amon}. With the rapid progress of observational cosmology, such discrepancies have been more significant, suggesting potential limitations of the $\Lambda$CDM model and motivating the search for new physics beyond the standard cosmological problem. 

As the fundamental nature and origin of DE remain elusive, cosmologists have proposed a broad spectrum of theoretical models to interpret this cosmic phenomenon. In general, the proposals are classified into two categories: DE model \cite{demodel,demodel1} and modified gravity theories \cite{gr,gr1}. The former class retains the framework of Einstein's general relativity but introduces additional components, such as scalar fields, exotic fields or dynamical vacuum energy to account for the observed accelerated expansion. Prominent examples such as quintessence, k - essence and phantom field models attribute cosmic acceleration to a time varying energy component with negative pressure. Dynamical DE models, such as quintessence scalar fields, have therefore been widely explored \cite{Copeland2006, Tsujikawa2013, Huterer2018}. The latter one postulates that general relativity may break down on cosmological scales and therefore modifies the Einstein - Hilbert action by introducing alternative gravitational terms motivated by deeper physical principles. Representative theories in this category include $f(R)$ gravity, Brans - Dicke theory and Gauss - Bonet modifications. They are all aiming to reproduce the effects of DE through geometric or curvature - based mechanisms rather than involving a separate energy field.  

In thermodynamics, every realistic fluid system inherently involves dissipative processes, which refer to the irreversible loss of energy due to internal friction and viscosity. These processes play a significant role in cosmological dynamics, as they influence how the Universe evolves over time. In cosmology, such dissipation manifests through bulk and shear viscosities within the stress–energy tensor of the cosmic fluid \cite{eckart,prisco}. While shear viscosity, which relates to the resistance against shape deformation, can usually be ignored on large cosmic scales due to the Universe’s isotropy and homogeneity. On the other hand, bulk viscosity, associated with resistance to volume changes during cosmic expansion, becomes crucial. The effects of bulk viscosity have been extensively explored in the frameworks of viscous dark matter (VDM) \cite{vdm,vdm1}, viscous dark energy (VDE) \cite{vde} - \cite{vde3}, and cosmic inflation. Interestingly, viscosity has been proposed as a potential factor behind the present acceleration of the Universe. Apart from influencing the Hubble expansion, viscosity also leads to additional entropy production, which heats both baryonic and DM components. Therefore, it influences the thermal history and structure formation of the Universe. 

Another line of investigation involves allowing DE to exchange energy with DM. Interaction terms of the form $Q \propto H \rho$ are consistent with observations and may alleviate the coincidence problem as well as the $H_0$ and $\sigma_8$ tensions \cite{Wang2016, Kumar2017, Yang2021, Poulin2023}. In addition, such interactions can emerge naturally in frameworks where the DM mass depends on the scalar field \cite{Amendola2000, Fardon2004}. A further intriguing theory that has arisen as a potential framework to provide fresh perspectives on this cosmic interplay is mass - varying DM. Here, we explore the mathematical models that describe the DE - DM interaction, concentrating on mass - varying DM. The theory of mass - varying DM postulates that the mass of DM particles, which is commonly represented by the symbol $(m)$, is a function of time rather than a constant, $(m = m(t))$. This theory presents a scalar field to characterize the variance in DM mass, which is commonly represented as $(\phi)$. For mass -varying DM, the Lagrangian can be written as: $
\mathcal{L}_{\text{mass-varying DM}} = \frac{1}{2}g^{\mu\nu}\partial_\mu\phi\partial_\nu\phi - V(\phi) + \mathcal{L}_{\text{DM}}(m(t), \Psi_{\text{DM}}),
$
where,
\(g^{\mu\nu}\) is the metric tensor. \(\partial_\mu\) represents the covariant derivative with respect to spacetime coordinate \(\mu\). \(V(\phi)\) is the potential energy associated with the scalar field \(\phi\). \(\Psi_{\text{DM}}\) represents the DM fields.
An effective coupling term that modulates the interaction intensity between the scalar field \(\phi\) and DE fields describes the interaction between mass - varying DM and DE.

On the geometric side, Lyra \cite{Lyra1951} proposed a modification of Riemannian geometry through a displacement vector field. Sen \cite{Sen1957} later formulated field equations in this setting, which reduce to general relativity when the displacement vector vanishes. Constant or time-varying displacement vectors can mimic a cosmological constant or stiff matter, offering novel mechanisms to model cosmic acceleration \cite{Halford1970, Pradhan2014, Biswas2020}. Recent work has extended scalar-tensor theories to Lyra geometry, yielding modified Friedmann equations with rich dynamics \cite{Bhamra2020, Singh2022}.

In this work, we investigate a cosmological framework that integrates Lyra geometry, a canonical scalar field and bulk viscous dark energy interacting with mass-varying dark matter. Through a detailed dynamical systems analysis, we show that the interplay of the Lyra displacement field and viscous effects. The model admits stable de Sitter like attractors driven by viscous dark energy and supports smooth transitions between matter and dark-energy–dominated epochs. Using Pantheon+ SNe Ia, BAO measurements and the SH0ES $H_0$ prior, we further constrain the model parameters via MCMC analysis, finding excellent agreement with current low-redshift observations. The small negative interaction parameter and the allowed viscosity values help reproduce the present expansion rate and may alleviate the $H_0$ tension. Overall, the combined dynamical and observational results indicate that viscous dark energy in the Lyra manifold is a viable extension of standard cosmology capable of providing a unified description of cosmic evolution.

The structure of this paper is outlined as follows. Section 2 deals with the formation of the mathematical model consisting of interacting viscous DE and mass-varying DM in the background of Lyra geometry. Section 3 addresses the dynamical analysis of the evaluated model through critical point studies. It also includes physical interpretations of the evaluated critical points and a stability analysis of such critical points. Section 4 focuses on the cosmological stability and instability through phase portrait analysis. In section 5, we present a Markov Chain Monte Carlo (MCMC) analysis using some recent observational datasets to validate our result observationally. Lastly, in section 6 we conclude the paper.

\section{Model Formulation: Interacting Viscous Dark Energy in Lyra Geometry}

In Lyra’s modification of Riemannian geometry \cite{Lyra1951}, the affine connection is altered by the introduction of a displacement vector field $\psi_\mu$ and corresponding metric tensor $g_{\mu\nu}$. Sen \cite{Sen1957} showed that the corresponding field equations can be written as
\begin{equation}
G_{\mu\nu} + \frac{3}{2}\psi_\mu \psi_\nu - \frac{3}{4}g_{\mu\nu}\psi_\alpha \psi^\alpha = T_{\mu\nu},
\label{eq:lyra_field_eq}
\end{equation}
where $G_{\mu\nu}$ is the Einstein tensor and $T_{\mu\nu}$ denotes the matter energy--momentum tensor. When $\psi_\mu=0$, the above reduces to Einstein’s field equations.

For a spatially flat Friedmann--Lemaître--Robertson--Walker (FLRW) spacetime
\begin{equation}
ds^2 = -dt^2 + a^2(t)\,(dr^2 + r^2 d\Omega^2).
\end{equation}
We assume $\psi_\mu=(\beta(t),0,0,0)$, where $\beta(t)$ is a function of cosmic time. The energy-momentum tensor consists of pressureless DM and DE described by a scalar field with bulk viscosity. Explicitly,
\begin{equation}
T_{\mu\nu} = (\rho_{\text{tot}} + p_{\text{tot}}) u_\mu u_\nu - p_{\text{tot}} g_{\mu\nu},
\end{equation}
where $\rho_{\text{tot}}$ is total density defined by $\rho_{\text{tot}}=\rho_m$(matter density)$+\rho_{\phi}$(energy density) and total pressure $p_{\text{tot}} = p_\phi^{\text{eff}}$. $u_\mu$ and $u_\nu $ indicate four velocity tensors.

The scalar field energy density and pressure are
\begin{align}
\rho_\phi &= \frac{1}{2}\dot{\phi}^2 + V(\phi), \\
p_\phi &= \frac{1}{2}\dot{\phi}^2 - V(\phi).
\end{align}

The presence of viscosity in DE fluid influences the dynamics of the Universe. The effective pressure at thermal equilibrium for a typical bulk viscous DE fluid can be expressed as 
\begin{equation}
p_\phi^{\text{eff}} = p_\phi - \zeta \theta,
\end{equation}
where $\zeta$ denotes bulk viscosity coefficient. The expansion scalar ($\theta$) can be written as $\theta = 3H = \frac{\dot{a}}{a}$, $H$ and $a$ stand for the Hubble parameter and scalar factor respectively. Therefore, the effective pressure can be restated as
\begin{equation}
p_\phi^{\text{eff}} = p_\phi - 3\zeta H.
\end{equation}
The negative contribution $-3\zeta H$ modifies the dilution rate of dark energy and can mimic phantom-like behavior when sufficiently large \cite{Brevik2017, Barbosa2021, Mostaghel2023}.

By considering these factors, the Lyra-modified Friedmann equations become
\begin{align}
3H^2 - \beta^2(t) &= \rho_m + \rho_\phi, \label{eq:friedmann1}\\
2\dot{H} + 3H^2 + \beta^2(t) &= -p_\phi^{\text{eff}}. \label{eq:friedmann2}
\end{align}
The terms involving $\beta^2(t)$ represent the contribution of Lyra’s displacement vector. For a constant $\beta$, this behaves as a cosmological constant; for a dynamical $\beta(t)$, it acts as an additional time-dependent dark sector component.

\subsection{Interaction in the Dark Sector}
In standard cosmology, dark matter and dark energy are assumed to evolve independently, each conserving its energy density separately. Here, we rely on this assumption and allow an explicit interaction term $Q$ that mediates energy transfer between the two sectors. The total conservation equation
\begin{equation}
\dot{\rho}_{\text{tot}} + 3H(\rho_{\text{tot}} + p_{\text{tot}}) = 0,
\end{equation}
is then split as
\begin{align}
\dot{\rho}_m + 3H \rho_m &= Q, \label{eq:dm_conservation}\\
\dot{\rho}_\phi + 3H(\rho_\phi + p_\phi^{\text{eff}}) &= -Q. \label{eq:de_conservation}
\end{align}
Therefore, a positive $Q$ corresponds to energy transfer from DE to DM, while a negative $Q$ represents the reverse. The functional form of $Q$ is not uniquely determined; commonly adopted forms include
\begin{equation}
Q_1 = 3\delta \dot{\phi} \rho_m, \qquad Q_2 = 3\delta \dot{\phi} \rho_\phi, \qquad Q_3= 3\delta \dot{\phi}\frac{M'_m(\phi)}{M_m(\phi)}\rho_m, \qquad Q = 3\delta \dot{\phi} \left\{\rho_m + \rho_\phi+\frac{M'_m(\phi)}{M_m(\phi)}\rho_m\right\},
\end{equation}
with $\delta$ a dimensionless coupling parameter constrained by observations \cite{Wang2016,Kumar2017,Yang2021} and $M_m(\phi)$ is the scalar field-dependent mass of the matter component. Such interactions can ease the cosmic coincidence problem and affect the late-time attractors of the system.

\subsection{Effective Dark Sector Dynamics}
Combining Eqs.~\eqref{eq:friedmann1}--\eqref{eq:de_conservation}, we see that three distinct ingredients govern cosmic dynamics in this framework:
\begin{enumerate}
    \item The displacement field $\beta(t)$ from Lyra geometry, which adds a geometric correction that can mimic a cosmological constant or stiff component depending on its evolution.
    \item Bulk viscous pressure in the scalar field sector, which alters the effective equation of state and can drive accelerated or phantom-like expansion.
    \item Interaction between DM and DE, parametrized by $Q$, which redistributes energy in the dark sector and modifies the background expansion and structure growth.
\end{enumerate}

This combined setup provides a richer phase space than standard quintessence or interacting models, and allows us to investigate whether the interplay between Lyra geometry, viscosity, and interaction can naturally generate late-time accelerated attractors consistent with observations.

\subsection{Autonomous System Representation}
To study the asymptotic dynamics, we introduce dimensionless variables normalized by the Hubble scale:
\begin{equation}
x^2 = \frac{\dot{\phi}^2}{6H^2}, \quad y^2 = \frac{V(\phi)}{3H^2}, \quad \Omega_{\beta} = \frac{\beta}{\sqrt{3}H^2}, \quad \Omega_{Viscous} = \frac{\zeta}{H}.
\end{equation}
The density parameters are
\begin{align}
\Omega_\phi &= x^2 + y^2, \\
\Omega_{\beta}^2 &= \Omega_{Lyra},\\
\Omega_m &= 1 - x^2 - y^2 - \Omega_{Lyra}.
\end{align}
In the above equations \( \Omega_{\phi} \), \( \Omega_{m} \) ,\( \Omega_{Viscous} \), \( \Omega_{\beta} \), and \( \Omega_{Lyra} \) represent the fractional densities due to DE, DM, viscous fluid, Lyra geometric displacement field, and auxiliary term, respectively.
The effective equation of state parameter is
\begin{equation}
w_{\text{eff}} = x^2 - y^2 + \Omega_{Lyra} - \Omega_{Viscous}.
\end{equation}

Assuming an exponential potential $V(\phi) = V_0 e^{-\gamma \phi}$, mass varying exponential coefficient of dark matter, $M_m(\phi)= M_0e^{\alpha \phi}$ and interaction $Q = 3\delta \dot{\phi} \left\{\rho_m + \rho_\phi+\frac{M'_m(\phi)}{M_m(\phi)}\rho_m\right\}$, the dynamical equations for $(x,y,\sqrt{\Omega_{Lyra}},\Omega_{Viscous})$ with respect to $N = \ln a$ become
\begin{eqnarray} \left. \begin{array}{llll}\label{model1}
\frac{dx}{dN} &=\sqrt{\frac{3}{2}}\frac{\Omega_{Viscous}}{x} -\sqrt{\frac{3}{2}}\,\delta\Big[1-\Omega_{Lyra}
   -\alpha\big(1-\Omega_{Lyra}-x^2-y^2\big)\Big] \\
   &\quad + \frac{3}{2}x\big(\Omega_{Lyra}-\Omega_{Viscous}+x^2-y^2-1\big)
   -\sqrt{\tfrac{3}{2}}\,\gamma y^2, \\
\frac{dy}{dN} &= y\Big[\sqrt{\tfrac{3}{2}}\,\gamma x
   + \tfrac{3}{2}\big(1+\Omega_{Lyra}-\Omega_{Viscous}+x^2-y^2\big)\Big], \\
\frac{d\Omega_{\beta}}{dN} &= \frac{3}{2}\Omega_{\beta}\Big[-1+\Omega_{Lyra}-\Omega_{Viscous}+x^2-y^2\Big], \\
\frac{d\Omega_{Viscous}}{dN} &= \tfrac{3}{2}\,\Omega_{Viscous}
   \big(1+\Omega_{Lyra}-\Omega_{Viscous}+x^2-y^2\big).
\end{array}\right\}
\end{eqnarray}

This autonomous system allows us to identify critical points and determine their stability, thereby characterizing the possible cosmological scenarios emerging from viscous interacting DE in Lyra’s manifold.

\section{Dynamical Analysis}

To study the asymptotic behavior of the cosmological model under Lyra geometry with viscous effects, we perform a dynamical systems analysis. The system of autonomous equations governing the dimensionless variables is given by (\ref{model1}). The parameters \( \alpha \), \( \gamma \), and \( \delta \) denote coupling and potential slope constants.
\begin{align}
\frac{dx}{dN} &= \sqrt{\frac{3}{2}}\frac{\Omega_{Viscous}}{x} 
-\sqrt{\frac{3}{2}}\,\delta\Big[1-\Omega_{Lyra}
   -\alpha\big(1-\Omega_{\beta}^2-x^2-y^2\big)\Big]  \nonumber\\
   &\quad + \frac{3}{2}x\big(\Omega_{\beta}^2-\Omega_{Viscous}+x^2-y^2-1\big)
   -\sqrt{\tfrac{3}{2}}\,\gamma y^2, \label{eq:dx}\\[3pt]
\frac{dy}{dN} &= y\Big[\sqrt{\tfrac{3}{2}}\,\gamma x
   + \tfrac{3}{2}\big(1+\Omega_{\beta}^2-\Omega_{Viscous}+x^2-y^2\big)\Big], \label{eq:dy}\\[3pt]
\frac{d\Omega_{\beta}}{dN} &= \frac{3}{2}\Omega_{\beta}\Big[-1+\Omega_{\beta}^2-\Omega_{Viscous}+x^2-y^2\Big], \label{eq:dbeta}\\[3pt]
\frac{d\Omega_{Viscous}}{dN} &= \tfrac{3}{2}\,\Omega_{Viscous}
   \big(1+\Omega_{\beta}^2-\Omega_{Viscous}+x^2-y^2\big). \label{eq:dvisc}
\end{align}

\subsection{Critical Points of the Dynamical System}

We assume that the Lyra term satisfies 
\(\Omega_{Lyra}\equiv \Omega_{\beta}^2\), which ensures algebraic consistency among the equations.  
The last two differential equations can then be written as
\begin{align}
\frac{d\Omega_{\beta}}{dN} &= \tfrac{3}{2}\,\Omega_{\beta}\,F,
\qquad F=-1+\Omega_{\beta}^2-\Omega_{Viscous}+x^2-y^2,\\
\frac{d\Omega_{Viscous}}{dN} &= \tfrac{3}{2}\,\Omega_{Viscous}\,G,
\qquad G=1+\Omega_{\beta}^2-\Omega_{Viscous}+x^2-y^2.
\end{align}
At any equilibrium, each prefactor or bracket must vanish.  
Hence, a fixed point must satisfy
\[
\Omega_{\beta}=0\ \text{or}\ F=0,\qquad 
\Omega_{Viscous}=0\ \text{or}\ G=0.
\]
Since \(G-F=2\), both \(F=0\) and \(G=0\) cannot hold simultaneously. Below we list all admissible families of equilibrium points under this constraint.

\subsubsection{Family I: \(\Omega_{Viscous}=0\) (vanishing viscous component)}

In this case \(G=1+\Omega_{\beta}^2+x^2-y^2\).  
The conditions \(d\Omega_{\beta}/dN=0\) and \(dy/dN=0\) give
\begin{align}
&\Omega_{\beta}=0 \quad\text{or}\quad 
F=-1+\Omega_{\beta}^2+x^2-y^2=0,\\
&y=0 \quad\text{or}\quad 
\sqrt{\tfrac{3}{2}}\gamma x + \tfrac{3}{2}G=0.
\end{align}
Finally, \(dx/dN=0\) reduces to
\begin{equation}
0 = -\sqrt{\tfrac{3}{2}}\,\delta\Big[1-\Omega_{\beta}^2
   -\alpha\big(1-\Omega_{\beta}^2-x^2-y^2\big)\Big]
   + \tfrac{3}{2}x(\Omega_{\beta}^2+x^2-y^2-1)
   -\sqrt{\tfrac{3}{2}}\,\gamma y^2. \label{eq:x_equil}
\end{equation}

The simultaneous solution of these algebraic equations defines the equilibrium points.

\paragraph{(a) Pure scalar-field solutions.}
Setting \(\Omega_{\beta}=0\) gives:
\begin{align}
0 &= -\sqrt{\tfrac{3}{2}}\,\delta\Big[1-\alpha(1-x^2-y^2)\Big]
     + \tfrac{3}{2}x(x^2-y^2-1)
     -\sqrt{\tfrac{3}{2}}\gamma y^2,\\
0 &= y\Big[\sqrt{\tfrac{3}{2}}\gamma x+\tfrac{3}{2}(1+x^2-y^2)\Big].
\end{align}

\paragraph{Explicit critical points for \(\Omega_{\beta}=\Omega_{Lyra}=0,\ \Omega_{Viscous}=0\)}

Define the auxiliary quantities
\[
\begin{aligned}
D &\equiv \alpha\delta - \gamma,\\[4pt]
\Sigma &\equiv \sqrt{\,D\big(\alpha\delta\gamma^2 + 12\delta - \gamma^3 + 6\gamma\big) + 9\,},\\[6pt]
\Xi &\equiv -3\,\Sigma - D\Big\{\delta\big(\alpha(\gamma^2-12)-6\big)
   +\gamma\big(\Sigma-\gamma^2+6\big)\Big\} + 9 .
\end{aligned}
\]

Assuming \(D\neq 0\) and the radicands below are nonnegative, the four algebraic critical points with
\(\Omega_{Lyra}=\Omega_{\beta}^2=0\) and \(\Omega_{Viscous}=0\) read (the two independent
\(\pm\) signs under the square roots produce four combinations):

\[
P_{(s,t)}:\quad
\begin{aligned}
x_{(s,t)} &= -\,\frac{s\,\Sigma + \alpha\delta\gamma - \gamma^2 - 3}
                {2\sqrt{6}\,D}, \\[8pt]
y_{(s,t)} &= -\,\frac{t\,
    \sqrt{\dfrac{ -3\,s\,\Sigma - D\Big[\delta\big(\alpha(\gamma^2-12)-6\big)
    +\gamma\big(s\,\Sigma-\gamma^2+6\big)\Big] + 9 }{D}}}
    {2\sqrt{3\alpha\delta - 3\gamma}},\\[8pt]
\Omega_{\beta\, (s,t)} &= 0,\qquad
\Omega_{Viscous\,(s,t)} = 0,
\end{aligned}
\]

where \(s,t\in\{+1,-1\}\). Concretely, the four points are
\[
P_{(+,+)},\; P_{(+,-)},\; P_{(-,+)},\; P_{(-,-)}
\]
obtained by taking all combinations of \(s=\pm1\) (the sign in front of \(\Sigma\) in \(x\))
and \(t=\pm1\) (the overall sign of the square-root in \(y\)).

\paragraph{Existence conditions}
For any of the points \(P_{(s,t)}\) to be admissible (real-valued and regular) the following must hold:
\begin{enumerate}
  \item \(D = \alpha\delta - \gamma \neq 0\) .
  \item \(\Sigma^2 = D(\alpha\delta\gamma^2 + 12\delta - \gamma^3 + 6\gamma) + 9 \ge 0\).
  \item The inner radicand for \(y\),
  \[
  R_y \equiv -3s\Sigma - D\Big[\delta\big(\alpha(\gamma^2-12)-6\big)
    +\gamma\big(s\Sigma-\gamma^2+6\big)\Big] + 9,
  \]
  satisfies \(\dfrac{R_y}{D}\ge 0\).
  \item The denominator \(3\alpha\delta - 3\gamma = 3(\alpha\delta - \gamma)=3D\) must be positive; equivalently
  the sign of \(D\) must be consistent with the sign choice in the inner radicand so that the quotient is nonnegative.
\end{enumerate}

\paragraph{Physical significance (interpretation of these four branches)}
\begin{itemize}
  \item All four points lie in the subspace where the Lyra displacement and viscous fraction vanish:
        \(\Omega_{Lyra}=\Omega_{\beta}^2=0,\ \Omega_{Viscous}=0\). Thus they are \emph{pure scalar-field} equilibria: the cosmic dynamics at these points is controlled solely by the scalar field (kinetic + potential balance encoded in \(x,y\)).
  \item The two branches with \(x\) given by the \(\mathbf{+}\Sigma\) formula (points \(P_{(+,\pm)}\)) typically correspond to one family of scalar-dominated solutions where the sign of \(y\) distinguishes expanding vs.\ contracting potential contributions (or physically the sign of the scalar field velocity/potential pairing). The explicit form of \(y\) shows dependence on \(\alpha,\delta,\gamma\) and encodes whether potential energy dominates (\(y\) large) or not.
  \item The two branches with \(x\) given by the \(\mathbf{-}\Sigma\) formula (points \(P_{(-,\pm)}\)) are the complementary scalar-field roots coming from the other algebraic branch of the quartic/cubic system. Depending on parameters they can represent:
    \begin{itemize}
      \item \emph{Kinetic-dominated} like behaviour (if \(|x|\) is large and \(y\) small) — typically unstable / early-time repellers, or
      \item \emph{Potential-dominated} (if \(y\) large relative to \(x\)) — candidates for late-time accelerated attractors.
    \end{itemize}
  \end{itemize}

In order to signify the early stages of universe, we can find a special case:
\begin{enumerate}[(i)]
\item \(P:(x,y,\Omega_{\beta},\Omega_{Viscous})=(\pm1,0,0,0)\),\\[2pt]
    which exist only if \(\delta=0\).  
    They correspond to a kinetic-energy dominated universe (\(w_{\rm eff}=1\)), typically unstable.
\end{enumerate}

\paragraph{(b) Mixed Lyra--scalar equilibria.}

\paragraph{Explicit form of critical points($P_{\beta}$)}
For \(\Omega_{Viscous}=0\)  we have
\(
\Omega_{\beta}^2 = 1 - x^2 + y^2.
\)
Using the given model (\ref{model1}), we can have
\[
x = -\frac{3}{\sqrt{3/2}\,\gamma} = -\frac{\sqrt{6}}{\gamma}.
\]
The second branch (with \(x=-\sqrt{6}/\gamma\)) yields an explicit closed-form solution. Substituting
\(x_c = -\sqrt{6}/\gamma\) into the \(x\)-equilibrium equation and using
\(\Omega_{Viscous}=0\) and \(\Omega_{Lyra}=\Omega_{\beta}^2\) gives the following exact coordinates:

\[
P_{\beta}:\quad
\begin{aligned}
x_c &= -\frac{\sqrt{6}}{\gamma},\\[6pt]
y_c &= \pm\,\frac{\sqrt{6}\,\sqrt{-\dfrac{\delta}{\,2\alpha\delta-\delta+\gamma\,}}}{\gamma}
    \;=\;\pm\frac{\sqrt{6}\,\sqrt{-\delta/(2\alpha\delta-\delta+\gamma)}}{\gamma},\\[8pt]
\Omega_{\beta c} &= \pm\frac{1}{\gamma}\,
\sqrt{ \dfrac{-12\alpha\delta + \gamma^2(2\alpha\delta-\delta+\gamma) - 6\gamma}
               {\,2\alpha\delta-\delta+\gamma\,} }\,,\qquad
\Omega_{Viscous\,c}=0.
\end{aligned}
\]

\textbf{Existence conditions} \begin{enumerate}
    \item \(
2\alpha\delta-\delta+\gamma < 0 \quad(\text{if }\delta>0),
\)

    \item \(
-12\alpha\delta + \gamma^2(2\alpha\delta-\delta+\gamma) - 6\gamma \ge 0.
\)
\end{enumerate}
Both conditions must hold simultaneously for a real-valued \(P_{\beta}\). (If \(\delta\) changes sign,
the inequality signs must be interpreted accordingly.)

\paragraph{Cosmological Significance}

The critical point $P_{\beta}$ represents a dynamical balance between the Lyra geometric modification and the scalar field evolution in the cosmological phase space. The absence of viscous contribution ($\Omega_{Viscous,c}=0$) suggests that this point corresponds to a stage where the universe behaves as an effective perfect fluid dominated by the scalar field and Lyra manifold contribution.

Physically, the term $\Omega_{\beta}$ measures the influence of the Lyra displacement vector on the cosmic dynamics, while the variables $(x, y)$ represent the kinetic and potential energy components of the scalar field, respectively. The signs of $y_c$ and $\Omega_{\beta c}$ correspond to distinct evolutionary branches:

\begin{itemize}
    \item The positive branch $(x_c, +y_c, +\Omega_{\beta c})$ is typically associated with an \emph{accelerating phase} of the universe, where the potential energy dominates and leads to a quintessence-like behavior.
    \item The negative branch $(x_c, -y_c, -\Omega_{\beta c})$ corresponds to a \emph{contracting or decelerating phase}, where the kinetic contribution of the scalar field dominates, possibly mimicking a stiff-fluid or matter-like regime.
\end{itemize}

The branch $P_{\beta}$ thus describes a cosmological configuration where:
\begin{itemize}
    \item The Lyra displacement field induces an additional geometric potential that modifies the late-time cosmic dynamics.
    \item The absence of viscous effects indicates that dissipation is negligible, corresponding to a nearly adiabatic universe.
    \item Depending on the choice of the coupling parameters $(\alpha,\delta,\gamma)$, the point $P_{\beta}$ can represent either a stable accelerated attractor (dark-energy–like behavior) or a transient scaling regime where the scalar field energy tracks the background matter content.
\end{itemize}

Hence, the critical branch $P_{\beta}$ provides a viable geometrically induced dark-energy configuration that emerges naturally within the Lyra manifold formulation and can describe both early and late-time cosmic acceleration depending on parameter tuning.

\subsubsection{Family II: \(\Omega_{\beta}=0\) (vanishing Lyra contribution)}

Now \(F=-1-\Omega_{Viscous}+x^2-y^2\) and \(G=1-\Omega_{Viscous}+x^2-y^2\).  
The conditions \(G=0\) or \(\Omega_{Viscous}=0\) yield:

\paragraph{(a) Pure scalar field}
\(\Omega_{\beta}=\Omega_{Viscous}=0\), identical to Family I(a).

\paragraph{(b) Viscous-scaling equilibria.}
For \(\Omega_{Viscous}\neq0\), set \(G=0\) giving
\begin{equation}
\Omega_{Viscous}=1+x^2-y^2.
\end{equation}
The \(y\)-equation then reads
\begin{equation}
y\Big[\sqrt{\tfrac{3}{2}}\gamma x + \tfrac{3}{2}G\Big]
 = y\Big[\sqrt{\tfrac{3}{2}}\gamma x + \tfrac{3}{2}(1+x^2-y^2-\Omega_{Viscous})\Big]=0,
\end{equation}
which is automatically satisfied since \(G=0\).  
The \(x\)-equation reduces to
\begin{equation}
0=\sqrt{\tfrac{3}{2}}\frac{\Omega_{Viscous}}{x}
 -\sqrt{\tfrac{3}{2}}\delta\Big[1-\alpha(1-x^2-y^2)\Big]
 +\tfrac{3}{2}x(x^2-y^2-\Omega_{Viscous})
 -\sqrt{\tfrac{3}{2}}\gamma y^2.
\end{equation}
Substituting \(\Omega_{Viscous}=1+x^2-y^2\) yields an algebraic equation for \(x,y\):
\begin{equation}
0=\sqrt{\tfrac{3}{2}}\frac{1+x^2-y^2}{x}
 -\sqrt{\tfrac{3}{2}}\delta\Big[1-\alpha(1-x^2-y^2)\Big]
 -\tfrac{3}{2}x(1)
 -\sqrt{\tfrac{3}{2}}\gamma y^2.
\end{equation}

Regularity of the original vector field requires \(x\neq0\) when \(\Omega_{Viscous}\neq0\)
(because \(f_1\) contains the term \(\propto \Omega_{Viscous}/x\)). Eliminating \(y^2\) via
\(\Omega_{Viscous}=1+x^2-y^2\) gives a single polynomial equation for \(x\) (in general cubic),
so the viscous-scaling equilibria are given implicitly by the real roots \(x_c\) of that polynomial,
with
\[
y_c^2 = 1 + x_c^2 - \Omega_{Viscous\,c},\qquad
\Omega_{Viscous\,c} = 1 + x_c^2 - y_c^2.
\]
In other words, the \(P_V\) points satisfy the above cubic-type algebraic equation and must be obtained
either by exact algebraic root formulas (lengthy closed-form expressions) or numerically for given parameters
\((\alpha,\delta,\gamma)\).

\paragraph{Cosmological Significance of the Family of Critical Points $P_V$}

The family of critical points $P_V$ corresponds to a regime in which the cosmic dynamics are dominated by the viscous component of the dark sector. These points are characterized by the coordinates
\[
P_V:\quad
\begin{aligned}
x_c &= x_V, \qquad
y_c = y_V,\\[3pt]
\Omega_{\beta\,c} &= 0, \qquad
\Omega_{\text{Viscous}\,c} = 1 + x_c^2 - y_c^2.
\end{aligned}
\]
Here, $x$ and $y$ represent the dimensionless kinetic and potential components of the scalar field, while $\Omega_{\text{Viscous}}$ quantifies the fractional energy density of the bulk viscous fluid. The vanishing of $\Omega_{\beta}$ indicates that the Lyra manifold correction is negligible at this stage, implying that the geometry behaves effectively as in standard general relativity but with viscous modifications.

\paragraph{Physical Interpretation}

At $P_V$, the dynamics of the universe are governed by an effective viscous fluid. The nonzero $\Omega_{\text{Viscous}\,c}$ represents a dominant bulk viscosity contribution, which acts as an effective negative pressure component. Such pressure naturally drives accelerated expansion and can mimic the behavior of dark energy. The scalar field variables $(x_c, y_c)$ contribute only as subdominant corrections, modulating the effective equation of state.

\paragraph{Effective Equation of State and Cosmic Evolution}

At the critical point $P_V$, the total effective equation of state parameter can be expressed as
\[
w_{\text{eff}} = x_c^2 - y_c^2 - \Omega_{\text{Viscous}\,c}.
\]
Using the constraint $\Omega_{\text{Viscous}\,c}=1+x_c^2-y_c^2$, we obtain
\[
w_{\text{eff}} = -1,
\]
which clearly describes a \emph{de Sitter} phase of the universe. This indicates that the $P_V$ family corresponds to a cosmological constant–like state driven not by a true vacuum energy but by viscous dissipation within the cosmic fluid.

\paragraph{Cosmological Role}

The family $P_V$ therefore represents a class of late-time attractor solutions characterized by:
\begin{itemize}
    \item Dominance of the viscous fluid sector ($\Omega_{\text{Viscous}\,c}>0$),
    \item Negligible contribution from the Lyra geometric field ($\Omega_{\beta\,c}=0$),
    \item An effective equation of state $w_{\text{eff}}=-1$, consistent with dark-energy–dominated accelerated expansion.
\end{itemize}

Such solutions can describe the transition of the universe from a matter-dominated decelerating era to a viscous-fluid–driven accelerated epoch. The damping nature of viscosity also stabilizes perturbations in the late universe, supporting the interpretation of $P_V$ as a stable attractor corresponding to a \emph{viscous de Sitter} regime. This branch thus provides a purely hydrodynamic explanation of dark energy behavior, emerging naturally within the viscous Lyra cosmological framework.

\subsubsection{Summary of equilibrium families}

\begin{table}[H]
\centering
\renewcommand{\arraystretch}{1.3}
\setlength{\tabcolsep}{6pt}
\small
\begin{tabular}{l| p{4.5cm} |p{3.5cm} |p{4.5cm}}
\hline
\textbf{Label} & \textbf{Critical Point $(x_c,\,y_c,\,\Omega_{\beta c},\,\Omega_{\text{Viscous}\,c})$} 
& \textbf{Dominant Component} & \textbf{Typical Behavior} \\
\hline
$P_{1,2}$ 
& $(\pm 1,\,0,\,0,\,0)$ 
& Scalar (kinetic) 
& Early-time, unstable \\

$P_{(+,\pm)}$ 
& $(x_c,\,y_c,\,0,\,0)$ 
& Scalar dominance 
& Late-time, stable (de Sitter) or early-time expansion, unstable \\

$P_{(-,\pm)}$ 
& $(x_c,\,y_c,\,0,\,0)$ 
& Kinetic-dominated (if $|x|$ large, $y$ small) or potential-dominated (if $y$ large) 
& Early-time expansion, unstable or late-time stable (de Sitter) \\

$P_{\beta}$ 
& $\Big(x_c,\,y_c,\,\sqrt{1 - x_c^2 + y_c^2},\,0\Big)$ 
& Lyra--scalar mix 
& Geometric scaling \\

$P_V$ 
& $\Big(x_c,\,y_c,\,0,\,1 + x_c^2 - y_c^2\Big)$ 
& Viscous fluid 
& Viscous scaling \\

\hline
\end{tabular}
\caption{Symbolic critical points (families) of the dynamical system.}
\label{tab:critical_points}
\end{table}

\subsubsection*{Physical interpretation}
\begin{itemize}
\item \(P_{1,2}\): Kinetic-energy dominated stiff-fluid era (\(w_{\rm eff}=1\)), representing early-time expansion before potential or geometric effects become important.
\item \(P_{(\pm,\pm))}\): 
\begin{enumerate}
    \item Potential-dominated or de Sitter phases (\(\omega_{\rm eff}\approx-1\)), typical late-time attractors.
    \item Matter-dominated phases$(\omega_{eff}<1)$ typically repellers
\end{enumerate}
\item \(P_{\beta}\): Mixed Lyra–scalar equilibria where the Lyra displacement field contributes effectively as a cosmological term.
\item \(P_V\): Viscous-scaling equilibria in which the effective bulk viscosity of the cosmic fluid balances the scalar-field energy, leading to possible accelerated but dissipative expansion.
\end{itemize}
\subsection{Stability Analysis: Eigenvalue Approach}

Recall the Jacobian entries (with \(s\equiv\sqrt{3/2}\)) derived earlier.
We denote the Jacobian evaluated at a critical point by \(J_c\) and its eigenvalues by
\(\{\lambda_i\}_{i=1}^4\).

\subsubsection*{(A) Kinetic points \(P_{1,2}=(\pm1,0,0,0)\)}

These points exist as equilibria only when the coupling parameter vanishes (i.e. \(\delta=0\)).

\[
\text{At }P_1=(1,0,0,0)\ (\delta=0):\qquad \text{Eigeenvalues }~~~
\lambda_1 = 3,\quad \lambda_2 = s\gamma + 3,\quad \lambda_3 = 0,\quad \lambda_4 = 3.
\]
\[
\text{At }P_2=(-1,0,0,0)\ (\delta=0):\qquad \text{Eigeenvalues }~~~
\lambda_1 = 3,\quad \lambda_2 = -s\gamma + 3,\quad \lambda_3 = 0,\quad \lambda_4 = 3.
\]
Because two eigenvalues equal \(3>0\), both \(P_{1,2}\) are generically \emph{unstable} (repellers).

\bigskip

\subsubsection*{(B) Scalar-only branches \(P_{(\pm,\pm)}\) with \(\Omega_{\beta}=\Omega_{Lyra}=0,\ \Omega_{Viscous}=0\)}

Let \((x_c,y_c)\) denote any root from the algebraic list \(P_{(+,+)},P_{(+,-)},P_{(-,+)},P_{(-,-)}\).
At such points \(\Omega_{\beta}=0,\ \Omega_{Viscous}=0\).

Two of the eigenvalues are given by the Lyra and viscous diagonal entries (they decouple):
\[
\begin{aligned}
\lambda_{\Omega_{\beta}}= \tfrac{3}{2}(-1 + x_c^2 - y_c^2),~~~~
\lambda_{\Omega_{Viscous}}= \tfrac{3}{2}(1 + x_c^2 - y_c^2).
\end{aligned}
\]
The remaining two eigenvalues come from the upper-left \(2\times2\) block acting on \((x,y)\):
\[
M_{2\times2} =
\begin{pmatrix}
A & B\\[4pt]
C & D
\end{pmatrix}_{\!c},
\]
where (evaluated at \(\Omega_{\beta}=\Omega_{Viscous}=0\))
\[
\begin{aligned}
A &\equiv \partial_x f_1\Big|_c
= -s\frac{0}{x_c^2} + s\delta(-2\alpha x_c) + \tfrac{3}{2}(x_c^2-y_c^2-1) + 3x_c^2,\\[4pt]
B &\equiv \partial_y f_1\Big|_c
= -s\delta(-2\alpha y_c) + \tfrac{3}{2}x_c(-2y_c) - 2s\gamma y_c,\\[4pt]
C &\equiv \partial_x f_2\Big|_c
= y_c\big(s\gamma + 3x_c\big),\\[4pt]
D &\equiv \partial_y f_2\Big|_c
= s\gamma x_c + \tfrac{3}{2}(1 + x_c^2 - y_c^2) - 3y_c^2 .
\end{aligned}
\]
The two eigenvalues \(\lambda_{1,2}\) in the \((x,y)\) sector are the roots of
\[
\lambda^2 - \tau\lambda + \Delta = 0,
\qquad
\tau = A + D,\quad \Delta = AD - BC.
\]
Thus
\[
\boxed{%
\lambda_{1,2} = \frac{\tau \pm \sqrt{\tau^2 - 4\Delta}}{2},
}
\]
with \(A,B,C,D\) as above evaluated at the chosen branch \((x_c,y_c)\).

\paragraph{Stability criterion for \(P_{(\pm,\pm)}\).}
\begin{itemize}
  \item If \(\lambda_{\Omega_{\beta}}<0,\ \lambda_{\Omega_{Viscous}}<0\) and \(\Re(\lambda_{1,2})<0\), the point is a \emph{stable attractor}.
  \item Mixed signs imply a \emph{saddle}.
  \item Presence of any positive real part implies \emph{instability}.
\end{itemize}

\bigskip

\subsubsection*{(C) Lyra–scalar mixed family \(P_{\beta}\) (with \(\Omega_{Viscous}=0\) and \(F=0\))}

Lets evaluate the Jacobian diagonal entries for the Lyra and viscous directions:
\[
\begin{aligned}
\lambda_{\Omega_{\beta}} = 3(1 - x_c^2 + y_c^2),~~~
\lambda_{\Omega_{Viscous}}= \tfrac{3}{2}\big(1 + (1 - x_c^2 + y_c^2) + x_c^2 - y_c^2\big) = 3.
\end{aligned}
\]
Therefore two eigenvalues are \(\lambda_{\Omega_{Viscous}}=3\) and \(\lambda_{\Omega_{\beta}}=3(1-x_c^2+y_c^2)\).

The remaining two eigenvalues come from the \(2\times2\) \((x,y)\) block evaluated at \(P_{\beta}\). Denoting those entries by \(A_\beta,B_\beta,C_\beta,D_\beta\) (forms analogous to the scalar-only case but with \(\Omega_{\beta}\neq0\) substituted). Their characteristic polynomial again reads
\[
\lambda^2 - \tau_\beta \lambda + \Delta_\beta = 0,
\qquad \tau_\beta = A_\beta + D_\beta,\;\; \Delta_\beta = A_\beta D_\beta - B_\beta C_\beta.
\]
Thus the two scalar eigenvalues are
\[
\lambda_{1,2} = \frac{\tau_\beta \pm \sqrt{\tau_\beta^2 - 4\Delta_\beta}}{2}.
\]

\begin{itemize}
  \item One eigenvalue equals \(3>0\) (the viscous direction), so \(P_{\beta}\) generically has an unstable direction unless non-generic cancellations occur; this tends to make \(P_{\beta}\) a saddle in many parameter regimes.
  \item Conclude: \(P_{\beta}\) is generically a \emph{saddle} because of the positive viscous eigenvalue, but specific parameter tuning may change this outcome (requiring further nonlinear/center-manifold analysis).
\end{itemize}

\bigskip

\subsubsection*{(D) Viscous-scaling family \(P_V\) (with \(\Omega_{\beta}=0\) and \(G=0\))}
Lets evaluate diagonal Jacobian entries for the \(\Omega\)-directions using \(\Omega_{\beta}=0\) and \(G=1+\Omega_{\beta}^2-\Omega_{Viscous}+x_c^2-y_c^2=0\):

\[
\lambda_{\Omega_{Viscous}} =-\tfrac{3}{2}(1 + x_c^2 - y_c^2) =-\tfrac{3}{2},~~\lambda_{\Omega_{\beta}}= \tfrac{3}{2}(-1 - \Omega_{Viscous\,c} + x_c^2 - y_c^2)
\]
Thus two eigenvalues are negative and simple: \(\lambda_{\Omega_{\beta}}=-3\) and \(\lambda_{\Omega_{Viscous}}=-3/2\).

The remaining two eigenvalues again come from the \((x,y)\) \(2\times2\) block evaluated at \(P_V\). Denote its entries by \(A_V,B_V,C_V,D_V\) (computable by substituting \(\Omega_{\beta}=0\) and \(\Omega_{Viscous}=1+x_c^2-y_c^2\) into the general derivative formulae). Their characteristic polynomial is
\[
\lambda^2 - \tau_V \lambda + \Delta_V = 0,
\qquad \tau_V = A_V + D_V,\quad \Delta_V = A_V D_V - B_V C_V,
\]
so the scalar-sector eigenvalues are
\[
\lambda_{1,2} = \frac{\tau_V \pm \sqrt{\tau_V^2 - 4\Delta_V}}{2}.
\]

\paragraph{Stability criterion for \(P_V\).}
\begin{itemize}
  \item Two eigenvalues are strictly negative (\(-3\) and \(-3/2\)). If \(\Re(\lambda_{1,2})<0\) as well, then all four eigenvalues have negative real parts and \(P_V\) is a \emph{stable node/spiral} (late-time attractor) — this is the common viscous-de Sitter attractor scenario.
  \item If one of \(\lambda_{1,2}\) has positive real part, \(P_V\) is a \emph{saddle}.
\end{itemize}

\section{Phase Portrait Analysis: Cosmological Stability and Instability}

In this section, we present the global phase portrait evolution and corresponding trajectories of the cosmological dynamical system under various regimes of the interaction and viscous parameters. Each portrait illustrates the evolution of the system in the $(x,y)$-plane and the corresponding cosmological density parameters $\Omega_{m}$, $\Omega_{Lyra}$, and $\Omega_{Viscous}$ as functions of the e-folding number $N$.  
The nature of the critical points and their stability are analyzed for each case using linear perturbation theory around the equilibrium points.

\subsection{Interaction-free and Non-viscous Dark energy Component}

\begin{figure}[H]
    \centering
    \includegraphics[width=8cm, height=6cm]{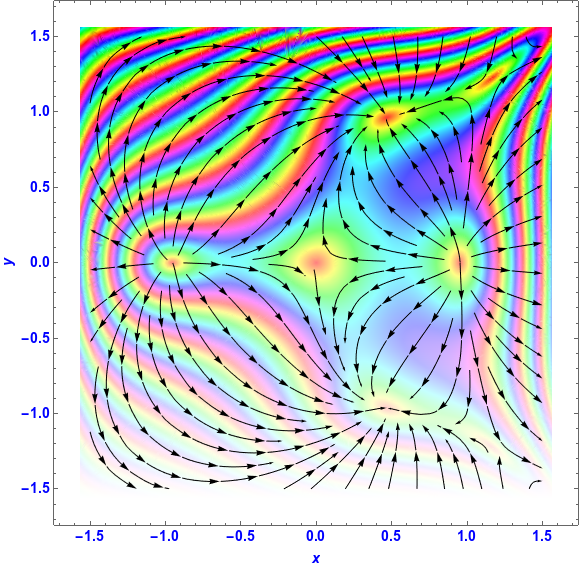}
    \includegraphics[width=8cm, height=6cm]{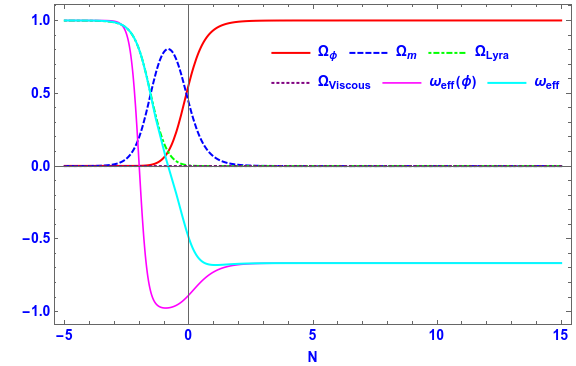}
    \caption{Phase portrait and corresponding evolution of cosmological parameters for $\gamma=-1, \alpha=1, \delta=0$ with initial values $x(0)=0.18$, $y(0)=0.73$, $\Omega_{\beta}(0)=0.065$, $\Omega_{viscous}(0)=0$. The system evolves toward a Lyra-dominated stable attractor.}
    \label{fig:null_null}
\end{figure}

For the case $\gamma = -1$, $\alpha = 1$, and $\delta = 0$, both interaction and viscous terms vanish. The phase portrait in Figure~\ref{fig:null_null} shows symmetric trajectories approaching the central attractor, indicating a stable configuration where the universe asymptotically evolves toward a Lyra-dominated de Sitter phase. The trajectories rapidly converge to the equilibrium state, and $\Omega_{Lyra} \to 1$ while $\Omega_{m}$ and $\Omega_{\phi}$ decay, implying a dark energy-displacement dominance due to presence of Lyra manifold geometry. In addition to this, we have two repeller points and one attractor point. For both repeller points, the speculated scalar potential revolves around zero. Again for $(x_c,y_c)\approx(0.49,0.64)$, which corresponds to a late-time accelerating phase of the universe with $\Omega_{\phi} \to 1$ while $\Omega_{m}$ and $\Omega_{Lyra}$ decay.

Moreover in the evolution part (right panel), we can observe that in the early universe $(N\approx-5)$ dominance of Lyra displacement is visible, after that gradually matter dominance begins almost from $N\approx-2.5$. Dark energy component grows at $N\approx-2$ as well as matter component decreases and gradually and it overtakes the matter component. At present $(N=0)$, graphs shows $\Omega_{\phi}=0.65,~\Omega_{m}=0.27$. In future if there is no interaction happened between non-viscous dark energy and mass-varying dark matter, dark energy will fully dominate as a result universe will expand exponentially.

\subsection{Interaction-free and Low Viscous Dark Energy}

\begin{figure}[H]
    \centering
    \includegraphics[width=8cm, height=6cm]{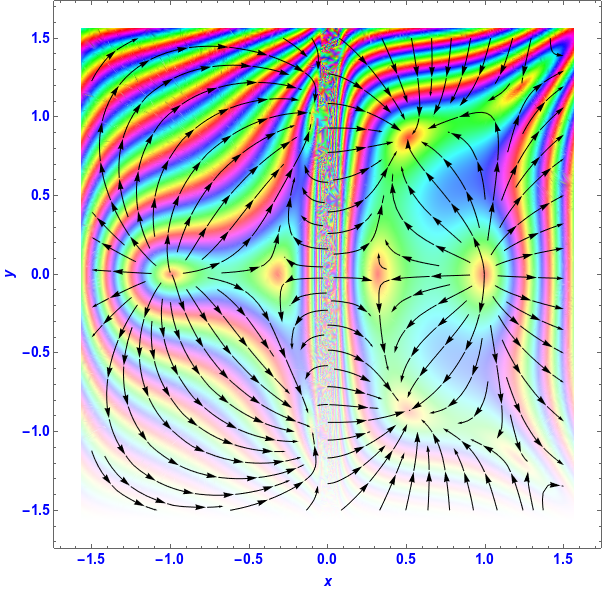}
    \includegraphics[width=8cm, height=6cm]{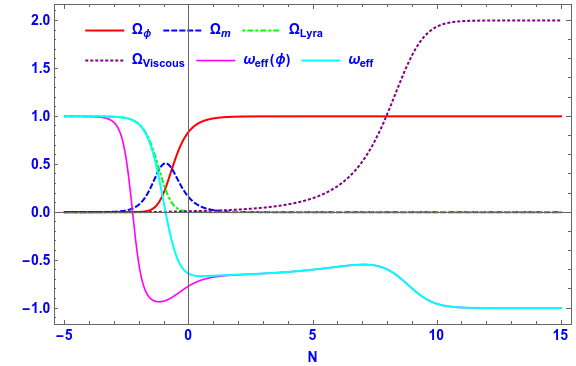}
    \caption{Phase portrait and cosmological evolution for $\gamma=-1, \alpha=1, \delta=0$, $x(0)=0.32$, $y(0)=0.86$, $\Omega_{\beta}(0)=0.065$, $\Omega_{viscous}(0)=0.01$. Weak viscosity leads to damped stable trajectories.}
    \label{fig:null_weak}
\end{figure}

When a small viscous component is introduced ($\Omega_{Viscous}(0)=0.01$), the phase portrait (Figure~\ref{fig:null_weak}) reveals slightly distorted trajectories, but the system still converges toward a stable attractor. The damping due to viscous dark energy slightly modifies the rate of convergence. There are four critical points in the viable region: two attractor and and two saddle points. 
Hence, the critical point with scalar potential $y_c>0$ i.e, $(x_c,y_c)\approx (0.56,0.82)$ remains stable, representing a universe evolving to a quasi–de Sitter attractor with small dissipative corrections.

In the evolution part (right panel), due to viscous property of dark energy dark matter is comparatively suppressed  and indicating a present situation where $\Omega_m=0.13$ and $\Omega_\phi=0.88$.

\subsection{Small Interaction and Non-viscous Dark Energy}

\begin{figure}[H]
    \centering
    \includegraphics[width=8cm, height=6cm]{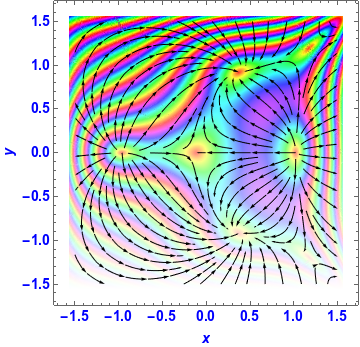}
    \includegraphics[width=8cm, height=6cm]{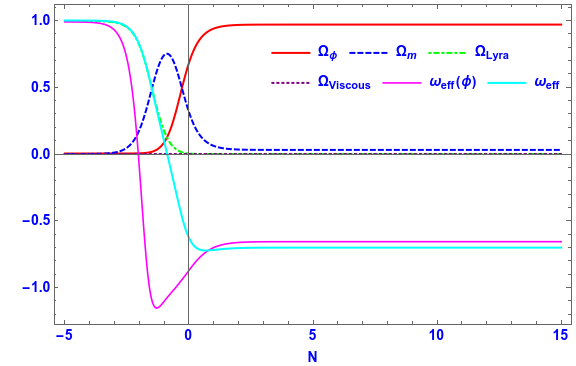}
    \caption{Phase portrait and parameter evolution for $\gamma=-1, \alpha=1, \delta=0.07$, $x(0)=0.16$, $y(0)=0.81$, $\Omega_{\beta}(0)=0.06$, $\Omega_{viscous}(0)=0$. Weak interaction shifts the stability basin, yielding one stable and one saddle point.}
    \label{fig:weak_null}
\end{figure}

Introducing a small interaction ($\delta = 0.07$) while keeping $\Omega_{Viscous}=0$ results in more dynamic trajectories (Figure~\ref{fig:weak_null}).  
Two distinct families of critical points appear — one unstable (saddle-like) near $(x, y) \approx (0, 0)$ and one stable node near $(x, y) \approx (0.4, 0.8)$. The interaction term shifts the source of attraction, allowing transient matter-dominated evolution before settling to the Lyra field attractor. Hence, the phase space consists of both transient (unstable) and asymptotically stable domains, representing a transition from matter-dominated era to energy dominance in the background of the Lyra manifold.

In addition, the cosmological evolution panel shows that dark matter increased in density just before the present era due to the small interaction introduced and ended up taking a higher value in the present situation. Again, the equation of state for dark energy $(\omega_{eff})$ will blow up beyond the phantom barrier in the near future.

\subsection{Small Interaction and Low Viscous Dark Energy}

\begin{figure}[H]
    \centering
    \includegraphics[width=8cm, height=6cm]{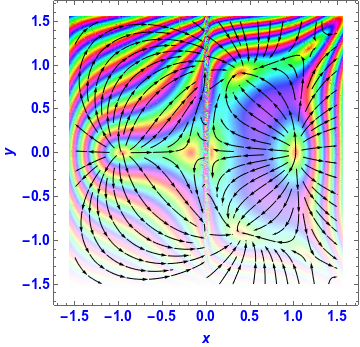}
    \includegraphics[width=8cm, height=6cm]{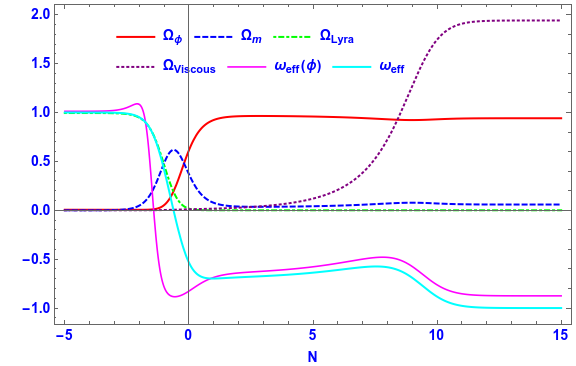}
    \caption{Phase portrait and cosmological evolution for $\gamma=-1, \alpha=1, \delta=0.07$, $x(0)=0.19$, $y(0)=0.76$, $\Omega_{d}(0)=0.13$, $\Omega_{viscous}(0)=0.01$. Both weak interaction and viscosity yield a globally stable focus-type attractor.}
    \label{fig:weak_weak}
\end{figure}

For small interaction and weak viscosity ($\delta=0.07$, $\Omega_{Viscous}(0)=0.01$), the portrait in Figure~\ref{fig:weak_weak} shows smoother spiraling trajectories approaching a stable focus. The combined damping from interaction and viscosity reduces oscillatory motion, producing a globally stable attractor. Hence, the system exhibits oscillatory convergence toward the de Sitter state, implying a stable late-time cosmological configuration.

\subsection{Substantial Interaction and High Viscous Dark Energy}

\begin{figure}[H]
    \centering
    \includegraphics[width=8cm, height=6cm]{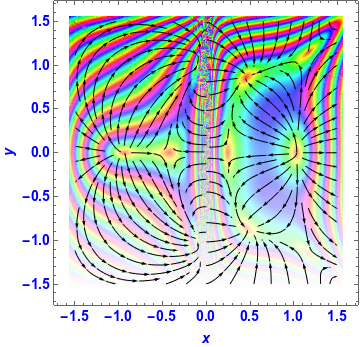}
    \includegraphics[width=8cm, height=6cm]{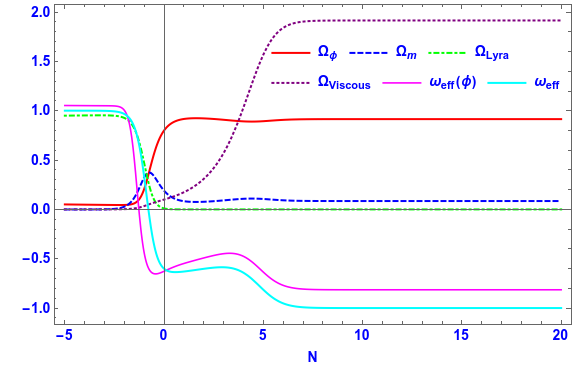}
    \caption{Phase portrait and corresponding evolution of cosmological parameters for $\gamma = -1$, $\alpha = 1$, $\delta = 0.07$, with assigned initial values $x(0)=0.38$, $y(0)=0.81$, $\Omega_{d}(0)=0.1$, and $\Omega_{viscous}(0)=0.1$. Strong coupling and viscosity enhance damping and alter the basin of attraction.}
    \label{fig:strong_strong}
\end{figure}

In the regime of strong coupling between the dark energy and highly mass-varying dark matter, along with significant bulk viscous effects of dark energy, the dynamical behavior becomes more complex. For the parameters $\gamma = -1$, $\alpha = 1$, $\delta = 0.07$ and the initial conditions $x(0) = 0.38$, $y(0) = 0.81$, $\Omega_{\beta}(0) = 0.1$, and $\Omega_{viscous}(0) = 0.1$, the system demonstrates the combined influence of strong viscous energy and interaction. The phase portrait in Figure~\ref{fig:strong_strong} exhibits multiple families of trajectories converging toward dominant attractors.  
Compared to the weak-viscosity cases, the trajectories are heavily damped and show minimal oscillations.  
This indicates that viscosity now plays a leading role in dissipating dynamical fluctuations, suppressing any transient oscillatory modes that might have existed in the weakly interacting cases.

From the cosmological parameter evolution (right panel), $\Omega_{\phi}$ gradually dominates as $N$ increases, while $\Omega_{m}$ and $\Omega_{Lyra}$ asymptotically approach small constant values. This behavior corresponds to a late-time accelerating phase where Lyra geometry drives the expansion but with residual energy contributions from matter and viscous effects. For this case, $\Omega_m$ does not vanishes due to sufficient interaction in the dark sector.

The strong dissipative regime ensures that any deviation from the equilibrium is rapidly suppressed, yielding a globally stable cosmological phase.  
Such a configuration represents a universe dominated by Lyra dark energy with minimal dynamical feedback from the interacting matter and viscous sectors.  
This regime thus corresponds to a robustly stable late-time de Sitter attractor, ensuring the asymptotic stability of the cosmic dynamics.

\subsection{Overall Summary of the Dynamical Regimes}

The five dynamical regimes presented in Figures~1--5 represent a systematic exploration of the cosmological system under varying interaction strengths ($\delta$) and viscous effects ($\Omega_{viscous}$). Each regime captures a distinct phase of the universe's evolution in the framework of Lyra geometry and presence of coupling viscous dark energy and mass-varying dark matter. The combined analysis of phase portraits and time-evolution plots reveals the transition from unstable early-time expansion to stable late-time acceleration as dissipation and interaction are gradually introduced.

The systematic progression from Regime~1 to Regime~5 demonstrates how the inclusion of interaction ($\delta$) and viscous effects enhances the dynamical stability of the cosmological model.  
While the null cases describe unstable or transitional early-time dynamics, the strongly interacting and viscous cases produce a stable de Sitter attractor consistent with observational late-time acceleration.  
Thus, viscosity and interaction play complementary roles: viscosity damps dynamical fluctuations, while interaction promotes geometric-matter energy exchange that drives the universe toward a Lyra–dominated equilibrium.

\begin{table}[H]
\centering
\renewcommand{\arraystretch}{1.2}
\small
\small{\begin{tabular}{p{5cm}| p{3cm} |p{2.5cm} |p{4.5cm}}
\hline
\textbf{Regime} & \textbf{Dominant Component} & \textbf{Stability Type} & \textbf{Cosmological Behavior}\\
\hline
Interaction-free \& Non-Viscous DE & Scalar (kinetic) & Unstable node & Early-time expansion \\
Interaction-free \& Low Viscous DE & Scalar + damping & Marginally stable & Onset of inflationary behavior \\
Small Interaction \& Non-Viscous DE & Lyra--scalar mix & Saddle-type & Intermediate scaling phase \\
Small Interaction \& Low Viscous DE & Viscous scaling & Stable spiral & Late-time accelerated attractor \\
Substantial Interaction \& High Viscous DE & Lyra dominated & Stable node & de Sitter late-time phase \\
\hline
\end{tabular}}
\caption{Summary of stability and cosmological interpretation for different interaction--viscosity regimes.}
\end{table}

\begin{figure}[]
	\centering
	\includegraphics[width=0.6\textwidth]{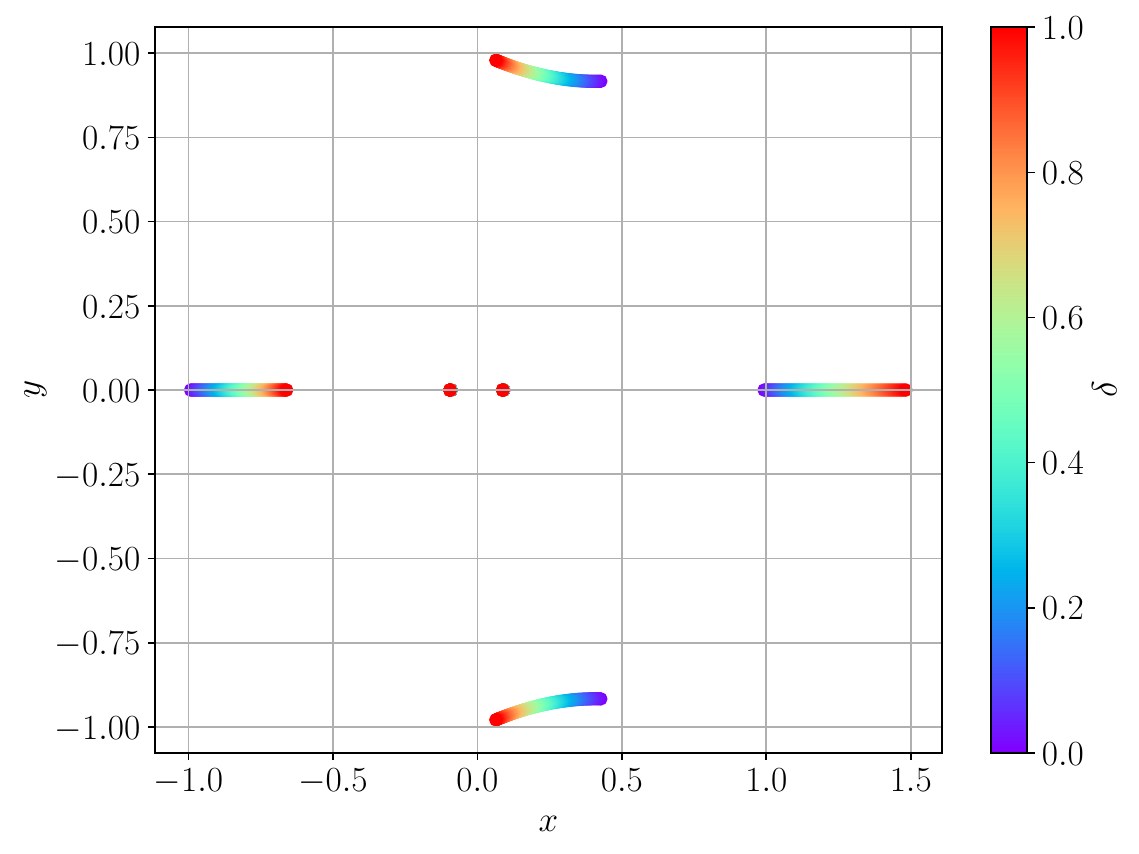}
	\caption{
		The figure shows how the stable points of the system change when the parameter $\delta$ varies, while keeping the other parameters fixed at $\alpha = 1.0$, $\gamma = -1.0$, $\Omega_{\beta 0} = 0.13$, and $\Omega_{\mathrm{vis}0} = 0.01$.}
	\label{fig:stable_points_delta}
\end{figure}

We extend our analysis by examining the behavior of the stable points under the variation of the viscous parameter $\delta$. 
Figure~\ref{fig:stable_points_delta} shows how these points change in the $(x, y)$ plane for different values of $\delta$, while keeping the other model parameters fixed at $\alpha = 1.0$, $\gamma = -1.0$, $\Omega_{\beta 0} = 0.13$, and $\Omega_{\mathrm{vis}0} = 0.01$. 
Different colors represent the values of $\delta$, which increases from violet to red, as indicated in the colorbar.

The red dots close to the origin $(0,0)$ indicate the stable points of the system among the total six stable points,  that do not move significantly with the change in $\delta$. 
However, as $\delta$ increases, the coordinates of the other four critical points shift along the plane, showing that the position and stability of the equilibrium states depend strongly on the viscous parameter. 
Two of the stable points lie near the $y$-axis but are located away from it, while the remaining stable points on the $x$-axis gradually move toward higher positive $x$ values as $\delta$ increases. 
This behavior suggests that stronger viscous effects lead to a more prominent response in the Universal dynamic and modify the overall stability structure of the model.

Overall, this analysis shows that the viscous parameter $\delta$ plays an important role in determining the evolution of the system. Even small changes in its value can modify the phase-space configuration and influence the late-time behavior of the cosmological model.

\section{Markov Chain Monte Carlo (MCMC) Analysis}

In order to constrain the free parameters of the model, a Markov Chain Monte Carlo (MCMC) analysis was carried out using a combination of recent observational datasets, including the Pantheon+ Type Ia supernovae sample \cite{Scolnic2022}, Baryon Acoustic Oscillation (BAO) data \cite{Eisenstein2005,Alam2017}, and the SH0ES measurement of the Hubble constant \cite{Riess2022}. The total likelihood was evaluated through the combined chi-square estimator,  defined as
\begin{equation}
	\chi^2_{\mathrm{tot}} = \chi^2_{\mathrm{SN}} + \chi^2_{\mathrm{BAO}} + \chi^2_{\mathrm{SH0ES}},
\end{equation}
where the contribution from the CMB shift parameter is not considered in the present computation to focus on low-redshift constraints.

For the supernova dataset, the distance modulus $\mu(z)$ was evaluated from the theoretical model prediction $\mu_{\mathrm{th}}(z)$, and the corresponding $\chi^2_{\mathrm{SN}}$ is obtained as
\begin{equation}
	\chi^2_{\mathrm{SN}} = \sum_{i}\left[\frac{\mu_{\mathrm{obs}}(z_i) - \mu_{\mathrm{th}}(z_i)}{\sigma_{\mu_i}}\right]^2,
\end{equation}
where $\mu_{\mathrm{obs}}$ and $\sigma_{\mu}$ denote the observed distance modulus and its uncertainty from the Pantheon+ compilation.

The contribution of BAO is incorporated through the inverse covariance matrix $C^{-1}$, following the standard procedure outlined in Refs.~\cite{Percival2010,Anderson2014}. The dilation scale $D_V(z)$ and the comoving angular diameter distance ere obtained from the numerically integrated Hubble function $H(z)$. The resulting $\chi^2_{\mathrm{BAO}}$ is expressed as
\begin{equation}
	\chi^2_{\mathrm{BAO}} = (\mathbf{T}_{\mathrm{th}} - \mathbf{T}_{\mathrm{obs}})^{\mathrm{T}} C^{-1} (\mathbf{T}_{\mathrm{th}} - \mathbf{T}_{\mathrm{obs}}),
\end{equation}
where $\mathbf{T}_{\mathrm{th}}$ and $\mathbf{T}_{\mathrm{obs}}$ are the theoretical and observed BAO ratios, respectively.

The local measurement of the Hubble constant from SH0ES provides a strong low-redshift prior, which is included via
\begin{equation}
	\chi^2_{\mathrm{SH0ES}} = \left(\frac{H_0 - 73.04}{1.04}\right)^2,
\end{equation}
with $H_0$ expressed in units of km\,s$^{-1}$\,Mpc$^{-1}$.

The MCMC sampling was performed using a uniform (flat) prior distribution for all free parameters within their physically motivated bounds. The posterior probability is therefore given by
\begin{equation}
	\mathcal{P}(\theta|D) \propto \exp\left[-\frac{1}{2}\chi^2_{\mathrm{tot}}(\theta)\right],
\end{equation}
where $\theta = (\alpha, \gamma, \log_{10}\delta, h_0, \log_{10}\chi_0, \log_{10}\psi_0, \log_{10}\Omega_{\beta 0}, \log_{10}\Omega_{\mathrm{vis}0})$ denotes the parameter vector.

\begin{figure}[H]
	\centering
	\includegraphics[width=0.9\textwidth]{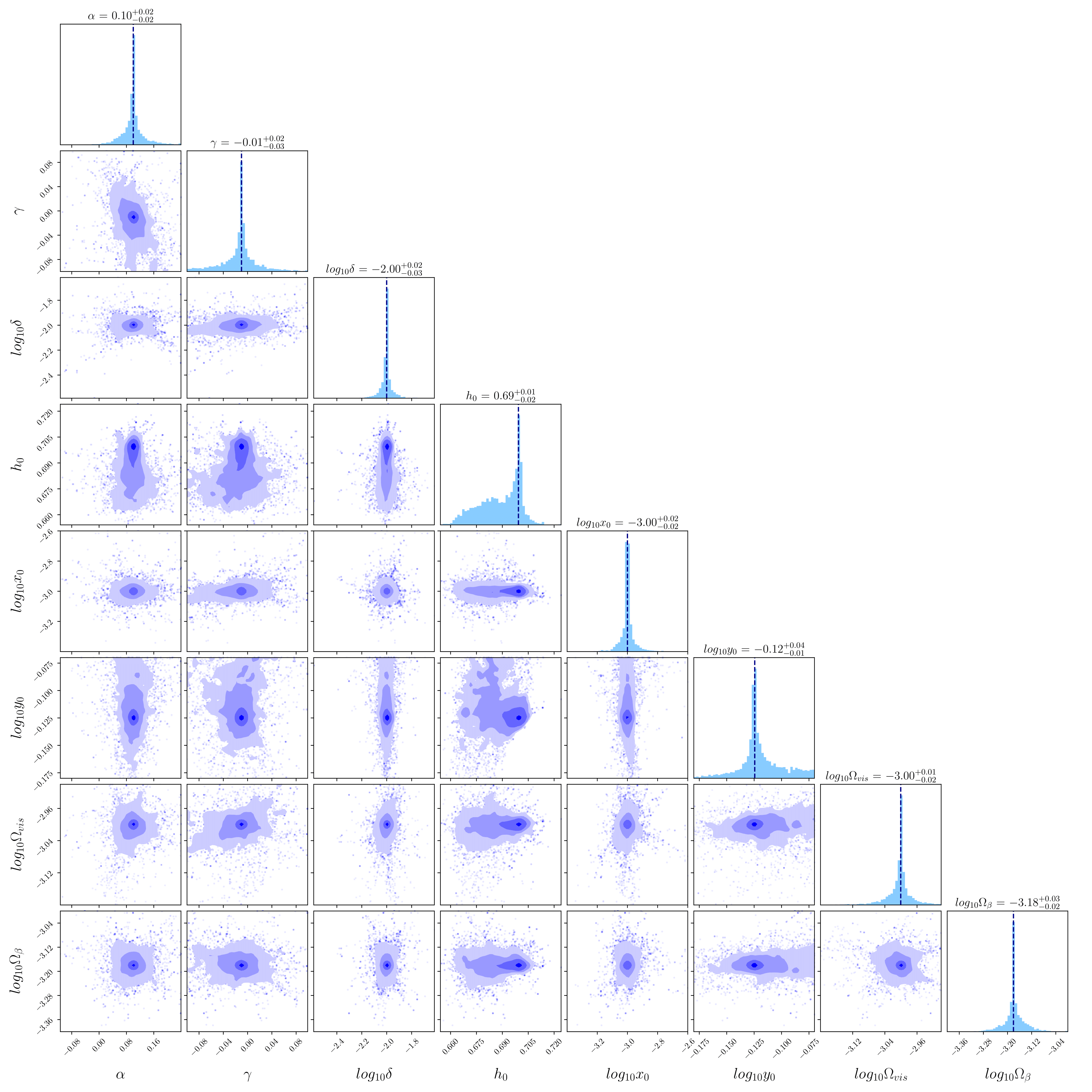}
	\caption{Corner plot showing the marginalized posterior distributions and 2D confidence regions (68\% and 95\%) for the model parameters. The diagonal panels represent the one-dimensional marginalized posteriors.}
	\label{fig:cornerplot}
\end{figure}

Figure~\ref{fig:cornerplot} shows the marginalized posterior distributions and the corresponding 2D confidence contours (68\%, 95\%) for all free parameters obtained from the MCMC chains. The parameters are found to converge smoothly with well-defined Gaussian-like posteriors, indicating good consistency among the datasets. The best-fit values (mean with $1\sigma$ uncertainty) of different parameters are tabulated in Table~\ref{tab:bestfit}.

\begin{table}[H]
	\centering
	\caption{Best-fit values of the cosmological parameters obtained from the MCMC analysis.}
	\begin{tabular}{cccccccc}
		\hline
		$\alpha$ & $\gamma$ & $\log_{10}\delta$ & $h_0$ & $\log_{10}\chi_0$ & $\log_{10}\psi_0$ & $\log_{10}\Omega_{\beta 0}$ & $\log_{10}\Omega_{\mathrm{vis}0}$ \\
		\hline
		$0.10^{+0.02}_{-0.02}$ & 
		$-0.01^{+0.02}_{-0.02}$ & 
		$-2.00^{+0.03}_{-0.03}$ & 
		$0.69^{+0.08}_{-0.08}$ & 
		$-3.00^{+0.02}_{-0.02}$ & 
		$-0.12^{+0.01}_{-0.01}$ & 
		$-3.00^{+0.08}_{-0.08}$ & 
		$-3.18^{+0.03}_{-0.03}$ \\
		\hline
	\end{tabular}
	\label{tab:bestfit}
\end{table}

\section{Summary and Conclusion}

In this study, we have carried out a detailed dynamical systems analysis of a cosmological model incorporating a Lyra geometric displacement field, a canonical scalar field and a bulk viscous dark energy component interacting with mass-varying dark matter. Our investigation demonstrates that the interplay of Lyra geometry and viscous effects provides rich dynamical behaviors capable of explaining both early- and late-time cosmic evolution within a unified framework.
The critical point analysis reveals several distinct equilibrium configurations representing different cosmological epochs. The kinetic-energy–dominated points $(P_{1,2})$ behave as early-time repeller states, consistent with stiff-fluid–like expansion scenarios ($w_{\rm eff}=1$) in the primordial universe \cite{copeland1998dynamics}. The scalar-field–dominated branches $(P_{(\pm,\pm)})$ contain both saddle and attractor solutions, with certain roots leading to potential-dominated accelerated expansion reminiscent of quintessence-driven de Sitter behavior \cite{ratra1988cosmological, wetterich1988cosmology}. These solutions align with the theoretical requirement that dark energy must dominate at late times, driving cosmic acceleration as supported by observational data from Supernovae Ia, CMB (Planck 2018), and BAO surveys \cite{aghanim2018planck, perlmutter1999measurements, riess1998observational}. The Lyra–scalar mixed branch ($P_{\beta}$) demonstrates how geometric modification through the Lyra displacement field acts as an effective cosmological term \cite{sen1971}. Although typically saddle-like, this branch contributes a viable transient stage connecting matter and dark energy phases. Most notably, the viscous-scaling family $(P_V)$ emerges as a dynamically robust late-time attractor, where viscous dark energy drives an effective equation of state $w_{\rm eff} \approx -1$, enabling de Sitter expansion without invoking an explicit cosmological constant. This connects well with the idea that dissipative processes such as bulk viscosity could act as a natural mechanism behind dark energy and cosmic acceleration \cite{brevik2005viscous, cataldo2005viscous}. The phase portrait analyses reinforce these findings, showing smooth transitions from matter-influenced epochs toward stable attractor solutions characterized by accelerated expansion. Weak viscosity and interaction maintain a de Sitter–like regime with mild damping, whereas strong viscous effects suppress dynamical oscillations rapidly, leading to a globally stable and observationally compatible accelerating universe. The model further indicates that residual matter and Lyra geometric contributions may persist at small but non-vanishing levels in the far future, consistent with dynamical dark energy expectations.

The obtained results highlight that Lyra geometry in conjunction with viscous dark energy provides a physically compelling framework capable of addressing key cosmological phenomena, including past deceleration, current acceleration, and asymptotic stability of the cosmic dynamics. Future work may incorporate perturbation analysis and observational constraints to refine parameter ranges and explore signatures distinguishable from the $\Lambda$CDM paradigm.

We further constrain the model parameters by performing a Markov Chain Monte Carlo (MCMC) analysis using a combined set of low-redshift cosmological observations—namely, the Pantheon+ Type Ia supernovae compilation \cite{Scolnic2022}, Baryon Acoustic Oscillation (BAO) measurements \cite{Eisenstein2005,Alam2017,Percival2010,Anderson2014} and the SH0ES Hubble constant prior \cite{Riess2022}. The likelihood is built through the total $\chi^2$ function combining these observed datasets and the numerical evolution of the model equations.

The posterior distributions of the model parameters are illustrated in Fig.~\ref{fig:cornerplot}. The best-fit values obtained (see Table~\ref{tab:bestfit}) are 
$\alpha = 0.10^{+0.02}_{-0.02}$, 
$\gamma = -0.01^{+0.02}_{-0.02}$, 
$\log_{10}\delta = -2.00^{+0.03}_{-0.03}$, 
$h_0 = 0.69^{+0.08}_{-0.08}$, 
$\log_{10}\chi_0 = -3.00^{+0.02}_{-0.02}$, 
$\log_{10}\psi_0 = -0.12^{+0.01}_{-0.01}$, 
$\log_{10}\Omega_{\beta 0} = -3.00^{+0.08}_{-0.08}$, and 
$\log_{10}\Omega_{\mathrm{vis}0} = -3.18^{+0.03}_{-0.03}$.
These values of the model parameters agree well with current cosmological observations. Our analysis shows that the model can describe the present expansion rate of the Universe and remains consistent with low-redshift observational data.

The interaction parameter $\gamma$ is small and negative, which indicates that there is a weak transfer of energy from dark matter to the viscous dark energy component. This energy exchange, along with the effect of bulk viscosity, helps the Universe accelerate at late times without needing a separate cosmological constant. The estimated Hubble parameter $h_0$ also matches well with the local SH0ES result, suggesting that including viscosity in the dark sector may help reduce the difference between local and global measurements of $H_0$. Overall, our results show that viscous dark energy models in the Lyra manifold give a stable and realistic description of the Universal evolution and are consistent with current observations.

\bibliographystyle{unsrt}

\end{document}